\documentstyle[prl,multicol,aps,amsmath,latexsym,float,epsfig]{revtex}

\def\gee        {\epsilon}

\def\go         {\omega}

\def\la         {\langle}
\def\ra         {\rangle}
\def\nk         {n{\bf k}}

\def\nukmq      {n_1\({\bf k-q}\)}

\def\dk         {\frac{d\,{\bf k}}{\(2 \pi\)^3}}
\def\dk1        {\frac{d\,{\bf k}_1}{\(2 \pi\)^3}}
\def\dq         {\frac{d\,{\bf q}}{\(2 \pi\)^3}}

\renewcommand{\[}{\left[}
\renewcommand{\]}{\right]}
\renewcommand{\(}{\left(}
\renewcommand{\)}{\right)}

\restylefloat{figure}

\begin{document}
\draft
\title{Quasiparticle Electronic structure of Copper in the GW approximation}
\author{
Andrea Marini, Giovanni Onida and Rodolfo Del Sole}
\address{
Istituto Nazionale per la Fisica della Materia e Dipartimento di Fisica
   dell'Universit\`a \\ di Roma ``Tor Vergata'',
   Via della Ricerca Scientifica, I--00133 Roma,
   Italy }
\date{\today}
\maketitle

\begin{abstract}
 We show that the results of photoemission and inverse
 photoemission experiments on bulk copper can be quantitatively
 described within band-structure theory, with no evidence of
 effects beyond the single-quasiparticle approximation.
 The well known discrepancies between the experimental
 bandstructure and the Kohn-Sham eigenvalues of Density
 Functional Theory are almost completely corrected by self--energy
 effects. Exchange--correlation contributions to the self--energy
 arising from 3s and 3p core levels are shown to be crucial.
\end{abstract}



\begin{multicols}{2}
\narrowtext


\noindent 
Experimental techniques for the determination of the electronic 
bandstructure of solids have made
considerable progress in recent years \cite{prexpcu}.
From the theoretical point of view, state--of--the--art 
calculations within many--body perturbation theory allow to 
obtain band energies in a rigorous way, i.e. as the poles of 
the one--particle Green's function $G$ \cite{gwrev}.  
To obtain  G, one needs the electron self-energy  $\Sigma$,
usually evaluated according to the so-called GW approximation,
derived by Hedin in 1965~\cite{hedin65}, which is 
based on an expansion in 
terms of the dynamically screened Coulomb interaction.
However, due to the high complexity and large computational 
requirements of ab--initio calculations of $\Sigma$, the
experimental bandstructures are often compared
with the results of (simpler) calculations 
performed within Density Functional Theory (DFT)\cite{DFT},
in the 
Local Density Approximation (LDA)\cite{LDA} or in the Generalized
Gradient Approximation (GGA)\cite{GGA}. The consequences of this 
approach (which can be set on a firm ground by considering
the exchange--correlation potential of the LDA or GGA as
an approximation to the self--energy operator) must 
however be considered with great care, particularly when
the system under study differs from those
--semiconductors and insulators-- for which the
approximation is usually made. 
A recent example is given by Ref. \cite{prexpcu}, 
where the measured Cu bandstructure is compared with 
state--of--the--art DFT-GGA results, finding discrepancies
which are both large and dependent on the considered band
and k--point. As a consequence of this result, the authors 
of Ref.  \cite{prexpcu} state that it is intriguing to find
such pronounced ``deviations from the band theory'', 
copper being much less correlated than metals with an open 3d shell
(e.g., Ni). 

In the present work, we demonstrate that copper does not, however,
deviate from quasiparticle band theory.
It only deviates from the simple
``rigid shift'' behavior of the self--energy commonly found
in the case of semiconductors and insulators.
For these, the application of the GW method to 
compute self--energy corrections  ontop of ab--initio DFT results 
has become a quite well-established and standard technique, 
giving energy levels generally in good agreement with
experiments, even for complicated systems like reconstructed
surfaces and clusters\cite{gwrev,surfclu}.
The gaps between empty and filled states generally increase 
by a substantial amount with respect to those
obtained in the Kohn-Sham (KS)
formulation of DFT, reaching agreement with experimental results.
The KS--DFT underestimation of the experimental gap 
is generally found to be weakly dependent on the particular band or k--point, 
although some semiconductor surface gaps show stronger QP corrections than bulk ones~\cite{olivia}.
This is the basis for the introduction of  the so--called ``scissors operator'',
often invoked in order to
correct the discrepancies by  rigidly shifting upwards the DFT empty bands, 
hence avoiding explicit self--energy calculations.
{\it A priori}, there is no reason why the ``scissors'' approach should 
also work for metals, 
where no gap exists between filled and empty states.

Unlike semiconductors, the case of metals 
has received limited attention so far. 
The  band width (i.e., the energy range of filled states) of 
simple metals has often been 
compared with that calculated for the homogeneous electron gas (jellium), 
finding discrepancies of a few tenths of eV\cite{Mahan}. 
On the other hand, the 
band structure of transition and noble metals 
cannot be approximated by that of jellium:   
valence electrons of d character play an important role
both in the electronic structure and in the 
optical properties of these metals.
Previous estimates of 
many--body corrections to the 3d Cu bandstructure, based on the  
self--energy of the  homogeneous electron gas, yielded only partial
agreement with the experimental data, suggesting that 
in the case of non--free electron metals 
screening effects (accounted for by the dielectric function) should be included in a more realistic
way\cite{nilsson}.
This is precisely the aim of the present work, where
a full, 
first--principles GW calculation of the quasiparticle (QP) bandstructure of bulk copper is
performed. 
The resulting QP energies
correct in a non trivial way the DFT-LDA results, and are in remarkably 
good agreement with
direct and inverse photoemission experiments.

Among transition metals, full quasiparticle 
calculations have been carried
out so far only for Ni\cite{aryas92}, while we are not aware of similar 
calculations for noble metals.
In the case of Ni, $GW$ yields a good description of
photoemission data, except for the $6$ eV satellite, which is due
to strong short--range correlations within the partially--filled d--shell. 
For copper, we expect band--theory to work better
than for transition metals, since d--shells are completely filled.
From the computational point of view, the very presence of strongly 
localized d--bands makes the ab--initio calculations based on plane waves (PW) 
and norm--conserving
pseudopotentials (NCPP) for noble metals much heavier than those for 
semiconductors. 

Let us first recall the relations between Kohn--Sham eigenvalues,
the poles of the one--particle Green's function G, and experimental band energies
measured in photoemission (PE) or inverse photoemission (IPE).
In DFT, KS eigenvalues  
cannot be identified with electron addition or removal 
energies, since there is no equivalent of Koopman's theorem.
Instead, the experimental bandstructure should be compared with the poles of the
Green's Function  $G\({\bf r},{\bf r}';\go\)$.
The latter are determined by an equation of the form:
\begin{multline}
 \[-\frac{\hbar^2}{2m}\triangle_{{\bf r}}+ V_{external}\({\bf r}\)+
  V_{Hartree}\({\bf r}\)\]\psi_{\nk}\({\bf r},\go\)+\\
 \int\,d{\bf r}'\,\Sigma\({\bf r},{\bf r}';\go\)\psi_{\nk}\({\bf r}',\go\)=
 E_{\nk}\(\go\)\psi_{\nk}\({\bf r},\go\),
    \label{eq:quasip}
\end{multline}
containing the  non-local, non-hermitian and frequency dependent
self-energy operator $\Sigma$.
The poles of G are the QP energies $\gee_{\nk}^{QP}$, the solutions of
$\gee_{\nk}^{QP}=E_{\nk}\(\gee_{\nk}^{QP}\)$.
Following  Ref. \cite{godby}, we separate the static, 
bare--exchange
part $\Sigma_x\({\bf r},{\bf r}'\)$ from $\Sigma_c\({\bf r},{\bf r}';\go\)$,
the energy--dependent correlation
contribution.
Eq. (\ref{eq:quasip}) is formally similar to Kohn-Sham equations which are
solved in the determination of the ground-state properties, 
where the local and energy independent exchange-correlation
potential $V_{xc}\({\bf r}\)$ has been substituted by $\Sigma$. 
Hence, KS eigenvalues can be considered as a zeroth-order approximation 
to the true QP energies, if the exchange-correlation (xc) potential of 
the DFT is seen as an approximation to the true self-energy operator $\Sigma$.
This suggests the possibility to look for a  first-order, 
perturbative solution of Eq. (\ref{eq:quasip}) with respect to
$\(\Sigma-V_{xc-LDA}\)$, where $V_{xc-LDA}$ is the xc potential of the LDA.
We have checked that the off--diagonal elements of $\(\Sigma-V_{xc-LDA}\)$
are two orders of magnitude smaller than the diagonal ones.
To determinate the QP energies $\epsilon^{QP}_{nk}$ starting from
the DFT-LDA values we have used an iterative procedure\cite{note}.

Our bandstructure calculation hence starts with a DFT-LDA calculation of the
ground-state properties, performed using norm-conserving pseudopotentials 
(PPs) and a plane waves basis.
The use of soft (Martins-Troullier\cite{MT}) PPs allows us to work at full convergence with a
reasonable kinetic energy cutoff (60 Rydbergs if the 3s and 3p
atomic states are frozen into the core, 160 Rydbergs when they are
explicitly included into the valence).
Details can be found in Ref. \cite{CUPRB}, where the 
non-trivial discrepancies between the LDA Kohn-Sham eigenvalues and the
experimental bandstructure have also been evidentiated.
In particular, we find substantial differences with respect to
the experiment for both  the d--bands width (3.70 instead of 3.17 eV)
and position (more than 0.5 eV upshifted in the DFT), in agreement
with previous calculations \cite{prexpcu,courths}.  
 
QP energies are then computed by evaluating $\Sigma$ as
GW \cite{hedin65},
so that the dynamically screened potential $W(\omega)$ (convolution
of the inverse dielectric function $\gee^{-1}\(\go\)$ 
with the bare Coulomb potential) is needed. 
Most GW calculations on semiconductor systems use a Plasmon-Pole Approximation 
(PPA) for  $W(\omega)$ ~\cite{PPA}, based on the observation that the Fourier components 
$\gee^{-1}_{{\bf G},{\bf G}'}\({\bf q};\go\)$
of the inverse dielectric function are generally peaked
functions of $\go$, and   can be approximated by a single pole.
Since the evaluation of $\Sigma_c$ involves an integration over the
energy, the fine details of the $\go$-dependence are not
critical, and the PPA turns out to work reasonably well for most applications.
However, in the case  of Cu the use of a PPA becomes more critical.
The presence of flat d--bands 2 eV below the Fermi level implies
the presence of strong transitions in $\gee^{-1}_{{\bf G},{\bf G}'}\({\bf q};\go\)$
spread over a large energy range.
For small values of ${\bf G}$ and ${\bf G}'$, the behaviour
of $\gee^{-1}_{{\bf G},{\bf G}'}\({\bf q};\go\)$
is often very different from that of a single-pole function, leading
to instabilities when determining the Plasmon-Pole parameters.
Hence, we have found it more convenient to avoid the PPA. 
Instead, we explicitly compute $\gee^{-1}_{{\bf G},{\bf G}'}\({\bf q};\go\)$
over a grid of about 200 frequencies from zero to $\sim$130 eV,
and  perform the energy integral numerically.
Another characteristic of metallic systems which leads to additional
difficulties in practical GW calculations is the discontinuity of the band
occupation crossing the Fermi surface.
The problem already shows up in the evaluation of the bare exchange 
matrix elements, which read:
\begin{multline}
\la \Sigma_x^{n {\bf k}} \ra \equiv \langle \nk | \Sigma_x\({\bf r},{\bf r}'\) | \nk\rangle = \\
 -\sum_{n_1}\int_{BZ}\dq \sum_{{\bf G}} 4 \pi \\
  \left|
 \frac{ \langle \nk | e^{i\({\bf q+ G}\)\cdot{\bf r}} | n_1{\bf k-q}\rangle}
 {{\bf q+G}}\right|^2 f_{\nukmq}.
 \label{eq:metal}
\end{multline}
Here $0\leq f_{n_{\bf k}}\leq 2$ is the occupation of band $n$ at point ${\bf
k}$, and its discontinuity at the Fermi Surface slows down the 
convergency of any numerical evaluation of Eq. (\ref{eq:metal}) on a finite
${\bf q}$-mesh, no matter how $\langle \nk | e^{i\({\bf q+ G}\)\cdot{\bf r}} |
n_1{\bf k-q}\rangle$ is smooth.
To overcome this difficulty, we have first computed the integral of the term
$f_{\nukmq}/\left|{\bf q+G}\right|^2$ over small regular volumes
centered around each ${\bf q}$-point used to evaluate Eq. (\ref{eq:metal}), and 
then used these values in the finite ${\bf q}$-sum.
A similar approach has been used in Ref.~\cite{olivia}.
This procedure yields results for $\la\Sigma^{\nk}_x-V^{\nk}_{xc-LDA}\ra$ computed with
19 ${\bf q}$-points in the IBZ  converged within 0.1 eV, while barely 
evaluating Eq. (\ref{eq:metal}) over the same set of ${\bf q}$-points yields an 
error of about  1.5 eV.

Normally (e.g., in GW calculations for semiconductors), the calculation of G and W 
to correct  the DFT valence bandstructure can
be performed by including only valence states, and fully 
neglecting the core states which have been frozen in the pseudopotential approach.
In copper, however, when $\Sigma$ is computed   neglecting the 3s and 3p 
atomic core states (which in the solid create two flat bands, at about 112 and
70 eV, respectively, below the Fermi level), the resulting QP corrections on the d-bands are
clearly unphysical: 
GW corrections move the highest occupied d--bands above
the  DFT-LDA Fermi level.
On the other hand, the
situation for s/p states (e.g. for the state $L_{2'}$) is much more reasonable, 
with correlation and exchange parts of the self-energy which cancel largely each
other (as in the case of semiconductors), and negative QP corrections of order of
the eV. 
\begin{figure}[H]
\begin{center}
\epsfig{figure=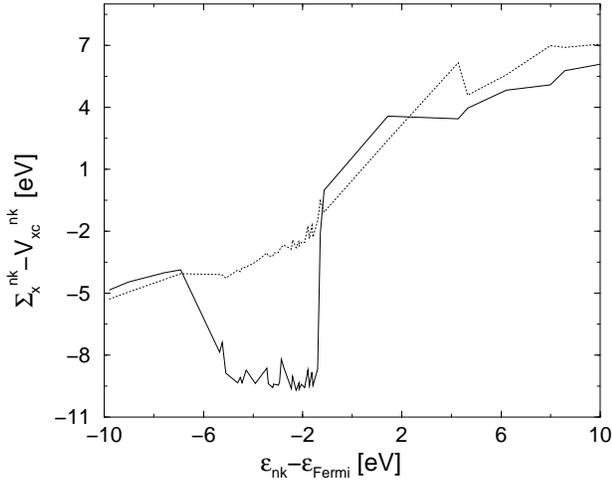,width=8cm}
\end{center}
\caption{\footnotesize{
Values of  $\la\Sigma^{\nk}_x-V^{\nk}_{xc}\ra$ for bulk copper,
plotted as a function of the
non-interacting energies $\gee_{\nk}$. The dotted line represents the results
obtained without
the contributions from the 3s and 3p  core states.}}
\label{fig:sx-vxc}
\end{figure}

The solution of this puzzling situation is provided by the role of
the above--mentioned 3s and 3p states, which, despite being 
well separated {\it in energy} from the 3d ones, have a large {\it spatial}
overlap with the latter.
As a consequence, non-negligible contributions to the self-energy are 
expected from the {\it  exchange} contributions 
between 3d and 3s/3p states.
These contributions can be included in our calculation 
in a straightforward way, by starting with a pseudopotential 
 which puts the whole $3^{rd}$ 
atomic shell into the valence (see Ref.\cite{CUPRB}).
Indicating with $G_0^{core}$ the non-interacting Green's function
containing also the  3s/3p levels,
the self-energy computed as $G_0^{core}W$ will yield the desired dynamical exchange 
between  3d bands and core bands composed of a bare exchange (static) contribution,
$\Sigma_x^{core}$, plus a correlation part, $\Sigma_c^{core}$.
Naturally, a larger plane-waves  cutoff (160 Ry) is needed for convergence.
As a result, we find large bare--exchange contributions from core levels,
leading to very different values of
$\la\Sigma^{\nk}_x-V^{\nk}_{xc-LDA}\ra$ for s/p and d states, as illustrated in 
Fig.~\ref{fig:sx-vxc}.
The role of core levels in the calculation of the bare exchange contributions (whose 
importance was already addressed for transition metals by Aryasetiawan and Gunnarsson~\cite{gwrev},
but estimated to be of the order of 1 eV) is hence crucial in the case of
copper~\cite{rohlfing}.
Moreover their effect is unexpected on the basis of
DFT--LDA calculations, where the band structure does not change
appreciably even if the {\it 3s/3p} orbitals are fully included in the valence~\cite{CUPRB}.
The effect of the core states is expected to be smaller on the
correlation part of $\Sigma\(\go\)$, since
the screened and bare Coulomb interaction, whose difference is involved
in $\Sigma_c^{core}$, are very similar to each other in the core region.
As a result, $\Sigma_c^{core}$ is small, of the order of $0.5$ eV for d--bands,
and close to zero for s/p--bands. 
The inclusion of the correlation part $\Sigma_{c}^{core}$ does not change
the widths of d-bands, that remain in good agreement with the 
experimental results, but improves considerably their
absolute positions\cite{notacorr}.
\noindent\begin{figure}[H]
\begin{center}
\epsfig{figure=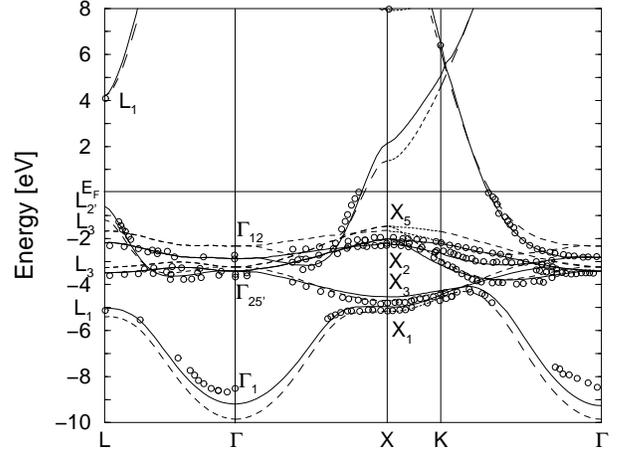,width=8cm}
\end{center}
\caption{\footnotesize{Full line: present GW results for the bulk copper bandstructure,
compared with the DFT--LDA results (dashed line), and with the experimental
data reported in reference {\protect\cite{courths}} (circles).  }}
\label{fig:bands}
\end{figure}

In Tab.~\ref{tab:gwres} we compare our calculations with selected
experimental data reported in Ref.~\cite{courths}.
In Fig.~\ref{fig:bands}  the full theoretical 
band structure is compared with experimental data: 
the agreement is remarkably 
good and the fact that the GW corrections cannot be reproduced
by any rigid shift of the LDA bands clearly appears. 
For instance, at the  L point 
the shifts change sign for different valence bands, with QP 
corrections ranging from $-0.29$ eV to $-0.61$ eV, to  $0.57$ eV 
for the three d--bands. 
As a conseguence, the $L_1-L_{2'}$ gap {\it decreases} after inclusion
of self--energy corrections (see also Tab.~\ref{tab:gwres}, last line), in
contrast with the usual behaviour occurring in semiconductors and insulators.
The other gaps at $L$ and $X$, instead, increase. All of them come to good
agreement with experimental data.
Unoccupied (conduction) bands at points L, X and K 
are also obtained in very good agreement with
the experimental data. These findings demonstrate that
the strong deviations from the  single quasiparticle 
band theory in copper, as suggested in Ref. \cite{prexpcu},
do not occur. QP band--theory in the $GW$ approximation, instead,
remarkably deviates from the DFT--LDA + scissors--operator approach, due to the
interplay of the different localization and correlation properties of d and s/p states.

In conclusion, we have carried out the first ab--initio 
quasi-particle calculation for a 
noble metal, copper. In order to do that, we had to overcome several 
problems, namely, the large number of plane waves needed to describe d orbitals, 
the calculation of k-space sums close to the Fermi surface, and the
inclusion of the exchange interaction between valence and core levels.
The single--quasiparticle bandstructure
computed within Hedin's $GW$ approximation for the
electron self--energy turns out to be in very good agreement with 
experiments. 
The GW method, originally devised to describe 
the {\it long-range} charge
oscillations \cite{hedin99}, is hence shown to yield a good description also
of copper, a system characterized by localized orbitals and 
short-range correlation effects. 
The corrections to the LDA Kohn--Sham eigenvalues for the d-bands 
are highly non--trivial, since even their {\it sign} turns out to be 
k-point and band--dependent, ruling out the validity of  
any ``scissors operator'' approximation.
Furthermore, exchange effects between d--electrons and 3s, 3p core 
states
play a  key role and cannot be neglected in the calculation 
of QP corrections. 
Finally, a full screening calculation of the 
self-energy has been carried out, avoiding the single-plasmon pole 
approximation. This allows a realistic treatment also of the imaginary part 
of the self energy (and therefore the calculation of QP lifetimes), 
which will be studied in a forthcoming work.

This work has been supported by
the INFM  PRA project ``1MESS'', MURST-COFIN 99 and
by the EU through the NANOPHASE Research Training
Network (Contract No. HPRN-CT-2000-00167).
We thank  Angel Rubio, Lucia Reining and Michele Cini for useful discussions, and
Conor Hogan for a critical reading.


\end{multicols}
\widetext


\newpage

\begin{table}
\begin{tabular}{lcccc}
 &   & DFT--LDA  &  $GW$ &  Experiment  \\ \hline
Positions & $\Gamma_{12}$ & $-2.27$ &  $-2.81$ &  $-2.78$   \\
of        & $X_5$         & $-1.40$ &  $-2.04$ &  $-2.01$   \\
{\it d}-bands   & $L_3$         & $-1.63$ &  $-2.24$ &  $-2.25$   \\ \hline
        & $\Gamma_{12}-\Gamma_{25'}$ & $0.91$ &  $0.60$ & $0.81$   \\
Widths  & $X_5-X_3$ &  $3.23$ &  $2.49$ &  $2.79$   \\
of      & $X_5-X_1$ &  $3.70$ &  $2.90$ &  $3.17$  \\
{\it d}-bands & $L_3-L_3$ &  $1.58$ &  $1.26$ &  $1.37$  \\
        & $L_3-L_1$ &  $3.72$ &  $2.83$ &  $2.91$  \\ \hline
Positions & $\Gamma_{1}$ & $-9.79$ &  $-9.24$ &  $-8.60$   \\
of {\it s/p}-bands & $L_{2'}$  & $-1.12$ &  $-0.57$ &  $-0.85$  \\ \hline
L-gap & $L_1-L_{2'}$ &  $5.40$ & $4.76$ &  $4.95$   \\
\end{tabular}
\caption{Theoretical band widths and band energies 
(LDA values + $GW$ corrections, in eV) 
for copper, at high-symmetry points. 
The striking agreement with the experimental results demonstrates
that copper is very well described within a single-quasiparticle
band theory, at the $GW$ level. The values in 
the last column are averages over several experiments, 
as reported in tab. 1 and 13 of ref.{\protect\cite{courths}}.
}
\label{tab:gwres}
\end{table}


\begin{references}
\bibitem{note}		Compared with the linearization method~\cite{godby} we 
			have found deviations smaller than $0.1$ eV.
\end{references}
\end{document}